\documentclass{article}
\title{Gr\"obner basis in Boolean rings is not polynomial-space}
\author{Mark van Hoeij\thanks{Supported by NSF grant 1319547} \\
Florida State University \\
Tallahassee, Fl 32306-3027, USA}
\usepackage{amsmath}
\usepackage{amsfonts}
\usepackage{amssymb}
\usepackage{amsthm}
\newtheorem{lemma}{Lemma}[section]

\newtheorem{thm}[lemma]{Theorem}
\theoremstyle{definition}

\newcommand{\eindebewijs}{\hfill$\Box$\par\medskip}
\begin{document}
\maketitle
\begin{abstract}
We give an example where the number of elements of a
Gr\"obner basis in a Boolean ring
is not polynomially bounded in terms of the bitsize and degrees of the input.
\end{abstract}
\section{Boolean Rings}
Let $R_n = \mathbb{F}_2[x_1,\ldots,x_n, y_1,\ldots,y_n, z_1,\ldots,z_n]$ where $\mathbb{F}_2 = \mathbb{Z}/(2)$.
If $S \subseteq R_n$ then $(S)$ denotes the ideal in $R_n$ generated by $S$, and
${\rm Sol}(S) \subseteq K^{3n}$ denotes the solution set of $S$, where $K$
is the algebraic closure of $\mathbb{F}_2$.
Let
\[ S_n := \{c^2 -c \,\, | \,\, c \in \{x_1,\ldots,x_n, y_1,\ldots,y_n, z_1,\ldots,z_n\} \}. \]
$R_n^{b} := R_n / (S_n)$ is a {\em Boolean ring}, which means $r^2 = r$ for all $r$ in $R_n^{b}$.
If $I$ is an ideal in $R_n$, then $S_n \subseteq I$ if and only if $I$ is radical and
${\rm Sol}(I) \subseteq \mathbb{F}_2^{3n}$.

Ideals in $R_n$ that contain $S_n$ are in 1-1 correspondence with ideals in $R_n^{b}$.
Thus, Gr\"obner basis in $R_n^{b}$ is equivalent to:  Gr\"obner basis in $R_n$ restricted to ideals that contain $S_n$.
We will show with an example that this is not polynomial-space.

\section{Example}
Let 
\[ L_n = \{ x_i y_i + x_i + y_i - z_i \,\,|\,\, i=1,\ldots,n\} \]
\[ T_n = \{ x_i z_i - x_i \,\,|\,\, i=1,\ldots,n \}  \bigcup \{ y_i z_i - y_i \,\,|\,\, i=1,\ldots,n \} \]
\[ P_n = \{ \prod_{i=1}^n c_i \,\,|\,\, c_1 \in \{x_1, y_1, z_1 \}, \, \ldots \, ,c_n \in \{x_n, y_n, z_n \}\}. \]
Let $G_n := S_n \bigcup L_n \bigcup T_n \bigcup P_n$ and $H_n = S_n \bigcup L_n \bigcup \{z_1 z_2 \cdots z_n\} \subseteq G_n$.
\begin{lemma} \label{lem1}
	$(H_n) = (G_n)$.
\end{lemma}
\noindent Proof: Both are radical so it suffices to show ${\rm Sol}(H_n) = {\rm Sol}(G_n)$.
If $x,y \in \mathbb{F}_2$ and $z=xy+x+y$ then $z=0 \Longleftrightarrow x=0 \wedge y=0$.
It follows that ${\rm Sol}(H_n)$ is the set of all $(x_1,y_1,z_1,\ldots,x_n,y_n,z_n)$
for which: $x_i, y_i \in \mathbb{F}_2$, $z_i = x_i y_i + x_i + y_i$, and $\exists_i \, x_i=y_i=0$.
These are solutions of $G_n \supseteq H_n$ as well. \eindebewijs

\begin{lemma}
If $n>1$ then $G_n$ is a reduced Gr\"obner basis of $(G_n)$ w.r.t. any admissible total-degree ordering.
\end{lemma}
Let ord be an admissible total-degree ordering. 
It is easy to check that $G_n$ is reduced when $n>1$, which means that for any $f,g \in G_n$ with $f \neq g$,
the ord-leading monomial of $f$ is not divisible by the ord-leading monomial of $g$. It remains to show that $G_n$ is a Gr\"obner basis. \\[5pt]
Proof: Let $B_n$ be the set of all monomials that are
not divisible by the leading monomial of an element of $G_n$
and let $V_n$ be the $\mathbb{F}_2$-vector space with basis $B_n$. The natural map $V_n \rightarrow R_n/(G_n)$ is always surjective
and is bijective if and only if $G_n$ is a Gr\"obner basis.
So it suffices to show that $V_n$ and $R_n/(G_n)$ have the same dimension.

Since $(G_n)$ is radical, dim $R_n/(G_n)$ is the number of solutions of $G_n$, which is $4^n-3^n$ (see the proof of Lemma~\ref{lem1}).
This equals the number of elements of
\[ B_n = \{ \prod_{i=1}^n c_i \,\,|\,\, c_1 \in \{1, x_1, y_1, z_1\}, \, \ldots \, ,c_n \in \{1, x_n, y_n, z_n\}\} - P_n. \]
\eindebewijs



\noindent To check the lemmas for a value of $n$, copy this in Maple:
\begin{verbatim}
n := 4;
vars := {seq(op({x[i], y[i], z[i]}), i=1..n)};
S := {seq(c^2-c, c=vars)};
L := {seq(x[i]*y[i] + x[i] + y[i] - z[i], i=1..n)};
H := S union L union {mul(z[i], i=1..n)};
G := Groebner[Basis](H, tdeg(op(vars)), characteristic = 2);
nops(H) = 4*n+1, nops(G) = 6*n + 3^n;
\end{verbatim}

\begin{lemma}
If $n>1$ then any total-degree ordered Gr\"obner basis of $(G_n)$ has at least $6n + 3^n$ elements.
\end{lemma}
\noindent Proof: $G_n$ has $|S_n| + |L_n| + |T_n| + |P_n| = 3n + n + 2n + 3^n = 6n + 3^n$ elements.
Gr\"obner bases are not unique, but any Gr\"obner basis has at least as many elements as a reduced Gr\"obner basis. \eindebewijs

\begin{thm}
Gr\"obner basis in Boolean rings is not polynomial-space.
\end{thm}
\noindent Proof: Let $n>1$. $H_n$ has $4n+1$ elements, whose degrees and bitsizes are bounded by a polynomial in $n$.
But a total-degree ordered Gr\"obner basis does not fit in polynomial-space
because it has at least $6n+3^n$ elements. \eindebewijs
\section{Comments}
The example was constructed by
converting $\exists_{i\in \{1,\ldots,n\}} \, x_i = y_i = 0$ to an ideal.
Converting this to a Gr\"obner basis increases the size exponentially.

Gr\"obner basis techniques are sometimes used for Boolean
expressions. Examples where this is beneficial can be found in Table~3 in~\cite{SAT}.

The previous version of this preprint used the phrase P-SPACE to denote
polynomial-space complexity because PSPACE is only for
decision procedures (e.g.: ideal membership in
Boolean rings is in PSPACE).
But P-SPACE is not standard terminology, so it is replaced by polynomial-space
in this version.

\end{document}